\documentclass[twoside,a4paper,11pt]{sea10}
\usepackage{graphicx}
\usepackage{hyperref}

\begin{document}
\pagenumbering{arabic}
\pagestyle{myheadings}
\thispagestyle{empty}
{\flushleft\includegraphics[width=\textwidth,bb=58 650 590 680]{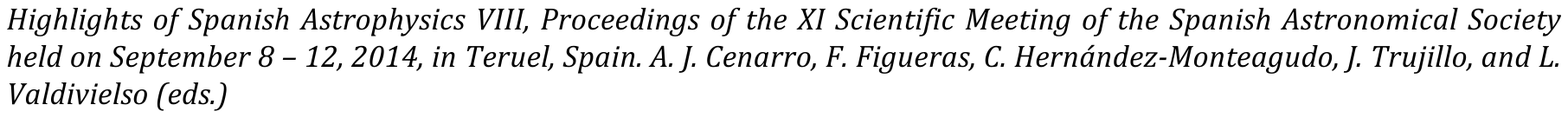}}
\vspace*{0.2cm}
\begin{flushleft}
{\bf {\LARGE
%
The Open Cluster Chemical Abundances from Spanish Observatories survey (OCCASO)
%
}\\
\vspace*{1cm}
%
R. Carrera$^{1,2}$,
L. Casamiquela$^{3}$, 
L. Balaguer-N\'u\~nez$^{3}$, 
C. Jordi$^{3}$, 
E. Pancino$^{4,5}$, 
C. Allende-Prieto$^{1,2}$, 
S. Blanco-Cuaresma$^{6,7}$, 
C. E. Mart\'{\i}nez-V\'azquez$^{1,2}$, 
S. Murabito$^{1,2}$, 
A. del Pino$^{1,2}$, 
A. Aparicio$^{1,2}$, 
C. Gallart$^{1,2}$, 
and 
A. Recio-Blanco$^{8}$
%
}\\
\vspace*{0.5cm}
%
$^{1}$
Instituto de Astrof\'{\i}sica de Canarias, La Laguna, Tenerife, Spain\\
$^{2}$
Departamento de Astrof\'{\i}sica, Universidad de La Laguna, Tenerife, Spain\\
$^{3}$
Dept. d'Astronomia i Meteorologia, Institut de Ci\'encies del Cosmos, Universitat de Barcelona (IEEC-UB), Barcelona, Spain\\
$^{4}$
Osservatorio Astronomico di Bologna, Bologna, Italy\\
$^{5}$
ASI Science Data Center, Frascati, Italy\\
$^{6}$
CNRS/Univ. Bordeaux, LAB, UMR 5804, Floirac, France\\
$^{7}$
Observatoire de Gen\`eve, Universit\'e de Gen\`eve, Sauverny, Switzerland\\
$^{8}$
Laboratoire Lagrange (UMR7293), Universit\'e de Nice Sophia Antipolis, CNRS, Observatoire de la Cote d'Azur, Nice, France
%
\end{flushleft}
%
\markboth{
OCCASO
}{ 
%
Carrera et al.
%
}
\thispagestyle{empty}
\vspace*{0.4cm}
\begin{minipage}[l]{0.09\textwidth}
\ 
\end{minipage}
\begin{minipage}[r]{0.9\textwidth}
\vspace{1cm}
\section*{Abstract}{\small
We present the motivation, design and current status of the Open Cluster 
Chemical Abundances from Spanish 
Observatories survey (OCCASO). Using the high resolution spectroscopic facilities available at 
Spanish observatories, OCCASO will derive chemical abundances in a sample of 20 
to 25 OCs older than 0.5 Gyr. This sample will be used to study in detail the 
formation and evolution of the Galactic disc using OCs as tracers.\normalsize}
\end{minipage}
%
%
%
\section{Introduction \label{intro}}

Stellar clusters are crucial in the study of a variety of topics including 
the star formation process, stellar nucleosynthesis and evolution, 
dynamical interaction among stars, or the assembly and evolution of 
galaxies. In particular, Open Clusters (OCs), which cover large ranges of ages 
and metallicities, have been widely used to constrain the formation and evolution of the 
Milky Way disc 
(e.g. \cite{2010A&A...511A..56P}).  This is because some 
of their features, such as ages or distances, are more accurately determined in 
comparison with field stars. They provide information about the chemical patterns and the existence of radial and vertical gradients or an age-metallicity relation. However, all these investigations are hampered by the fact that only a small fraction of clusters have been studied homogeneously.

Galactic surveys performed from the ground such as the APO Galactic Evolution 
Experiment 
(APOGEE; \cite{2013ApJ...777L...1F}), the Gaia-ESO Survey (GES; 
\cite{2012Msngr.147...25G}), or the GALactic Archaeology with HERMES (GALAH; \cite{2014IAUS..298..322A}) 
include OCs among their targets, providing radial velocities and 
chemical abundances. OCs are also sampled from the space by the Gaia (e.g. 
\cite{2001A&A...369..339P}) and Kepler missions. The first will provide accurate 
parallaxes, from which distances will be derived, and proper motions, and the 
second is providing accurate photometry.

\section{Survey design\label{sec2}}

The GES was designed to use the FLAMES (\cite{2002Msngr.110....1P}; GIRAFFE+UVES) capabilities at one of the 
VLT units in order to complement the Gaia mission. Among GES clusters observations are including 20-25 OCs older than 
0.5 Gyr. For them, GES is using the GIRAFFE fibers to derive radial velocities 
and abundances in stars at any evolutionary stage 
brighter than V$\sim$17 with a resolution lower than 20000. The six UVES 
fibers, which cover a wavelength range between 480 and 700 nm with a 
resolution of 47000, are being used to measure accurate radial velocities and 
detailed chemical abundances only in red clump stars. The UVES observations of old OCs 
have been designed to obtain an homogeneous sample of chemical abundances in 
order to study the Galactic disc. Using stars in the same evolutionary stage 
ensures the homogeneity of the sample.

Unfortunately, GES is sampling only the Southern hemisphere. However, several 
key OCs such as the most metal-rich, NGC~6791, and the 
oldest, Berkeley~17, together with several systems towards the Galactic anticenter 
or those observed by the Kepler mission are only visible from the North. APOGEE is 
the only survey that is sampling northern clusters in the \textit{H} band with a 
resolution of 22500. However, APOGEE is sampling OC stars at any evolutionary 
stage, like GES-GIRAFFE. Moreover, APOGEE is not observing a minimum of 
stars in each cluster. In fact, six or more cluster members have been analyzed  
only in 7 of the OCs
observed, those selected for calibration purposes (\cite{2013ApJ...777L...1F}).

The Open Cluster Chemical Abundances from Spanish Observatories survey (OCCASO) 
has 
been designed to complement from the North the GES-UVES observations of 
intermediate-age and old OCs in the 
South using the facilities available at the Spanish Observatories. OCCASO is 
being 
developed in parallel with GES. Like GES-UVES, OCCASO is observing a minimum of 
six red clump stars in a sample of 20 to 25 OCs older than 0.5 Gyr. Red clump 
stars are selected because they are easily identified even in the sparsely 
populated 
color-magnitude diagrams and their spectra are less line crowded and therefore, 
easier to analyze than that of brighter giants. Moreover, targeting objects in the 
same evolutionary stage avoids measuring anomalous abundances for some elements due to the effects of stellar evolution.  Therefore, 
at the end we will double the sample of OCs with homogeneous chemical abundance 
determinations. To ensure obtain abundances in the same scale than GES we are observing several 
stars in common and we are using some of the analysis methods also used in GES 
(see Sections~\ref{sec4} and \ref{sec5}). APOGEE is the only spectrograph with 
similar multi-object capabilities than UVES in the North but with lower 
resolution ($\sim$22500) and in the infrared. However, at Spanish 
observatories there are available several echelle high-resolution spectrographs 
with resolutions and wavelength coverage ranges similar or larger than UVES. In 
particular for this project we have selected CAFE@CAHA 2.2m 
(\cite{2013A&A...552A..31A}; 396$<\lambda<$950 nm, R$\sim$60000), 
FIES@NOT 2.5m (\cite{2014AN....335...41T}; 370$<\lambda<$750 nm, R$\sim$67000) 
and 
HERMES@Mercator 
1.2m (\cite{2011A&A...526A..69R}; 377$<\lambda<$900 nm, R$\sim$60000). Although 
only one star can be observed at once in each of them, the fact that we have 
distributed 
our observations among three different telescopes/instruments is allowing us to 
develop OCCASO in a timeline similar to GES. The brightest targets (V$\leq$12.5) 
are being observed 
with MERCATOR/HERMES, those stars with 12$\leq$V$\leq$14 are being observing 
with CAFE/CAHA 2.2m, and the faintest objects with FIES/NOT. For sake of 
homogeneity, we are observing stars in common in all telescopes.

\section{Observations, data reduction and radial velocity 
determination\label{sec3}}

OCCASO obtained 5 nights in each NOT and Mercator telescopes in semester 13B, and it 
was selected as a large program in the same telescopes from semester 14A which 
ensured 5 nights per semester and telescope during 4 semesters, till semester 
15A. 
Moreover, it is regularly obtaining Director Discretional and Spanish Guaranteed 
Times at the CAHA 2.2m telescope from semester 14A. Until November of 2014 we 
have completed a total of 44 observing nights with about 20\% of them lost by 
bad sky conditions. Moreover, the sky conditions of several of the observing 
nights were not good enough and we had to observe brighter stars than expected. 
In 
total, we have acquired 524 spectra for 87 stars belonging to 18 OCs. A minimum 
of 
6 stars have been observed in  11 clusters.

The FIES and HERMES spectrographs have dedicated pipelines which perform the bias 
subtraction, flat-field normalization, order trace and extraction, wavelength 
calibration, and order merge. For spectra acquired with CAFE we are using the 
pipeline developed by J. Maiz-Apellaniz. After this basic reduction, the spectra of the three 
telescopes are handled in the same way. Firstly, the sky emission lines are 
subtracted using a sky spectrum acquired during each run. Next, each 
spectrum is normalized by fitting the continuum with a polynomial function using 
DAOSPEC \cite{2008PASP..120.1332S}. The degree of the polynomial function changes 
from instrument to instrument. The telluric absorption lines are removed in the 
normalized spectra using a telluric star spectrum acquired in each run. The 
wavelength calibration is corrected of the heliocentric velocity. All the spectra of the same star and instrument are 
combined to reach the required signal-to-noise ratio. Finally, the radial 
velocity of each star is computed from the combined spectra using DAOSPEC and 
the same linelist used in the chemical abundance determination (see 
Section~\ref{sec4}).

\section{Chemical abundance determination\label{sec4}}

The OCCASO goal is to derive abundances for more than 20 chemical species, 
including light-elements (C, N), Fe-peak elements (Sc, V, Cr, Fe, Co, Ni), $\alpha$-elements (O, Mg, 
Si, Ca, Ti), s-process elements (Y, Z, Ba, La, Nd, Ce), proton-capture elements 
(Na, Al); the r-process element Eu; and Cu and Mn, elements with still unclear 
nucleosynthesis (e.g. \cite{2010ApJ...717..333C}). In order to ensure the 
reliability of the derived chemical abundances, these will be derived using 
different analysis techniques similar to what is being performed by GES.

The first approach is the classical one in which the abundances are derived from equivalent 
widths. In our case, the equivalent widths are measured with the 
automated wrapper DAOSPEC Option Optimizer (DOOp; \cite{2014A&A...562A..10C}) 
which automatically optimizes the best DAOSPEC inputs before 
determining the equivalent widths. In the next step, the abundances of 
individual spectral lines are derived using the WIDTH9 code developed by R. L. 
Kurucz implemented in GALA \cite{2013ApJ...766...78M}. Briefly, GALA determines 
the best model atmosphere by optimizing temperature, surface gravity, 
microturbulent velocity and metallicity, after rejecting the discrepant lines 
and it computes accurate internal errors for each atmospheric parameter and 
abundance. The GES linelist is used but extended to the blue by the 
\cite{2011A&A...535A..30C} linelist.

The second approach is the spectral synthesis in which the observed spectrum is 
compared with a library of 
synthetic spectra of known features in order to derive that which better 
reproduces the observed one. We are using three different tools to 
perform the spectral synthesis. FERRE \cite{2004AN....325..604A} identifies the 
model parameters that best reproduce the observations by means of an 
optimization algorithm that uses the chi-squared as metric. The MATrix 
Inversion 
for Spectral SynthEsis (MATISSE) algorithm (\cite{2006MNRAS.370..141R}) 
determines the atmosphere parameters 
on the basis of a linear combination of a grid of theoretical spectra. 
iSpec (\cite{2014A&A...569A.111B}) compares an observed spectrum with synthetic 
ones generated on the fly, using a least-squares algorithm but only in specific 
regions of 
the spectrum to minimize the computation time. 

Stellar parameters and chemical abundances for all observed stars until now have 
been derived with the classical method. We are working at this moment on 
deriving atmosphere parameters and abundances with the other methods. Derived 
stellar atmosphere parameters and Fe abundances for all stars observed until now will be 
released in the first data release scheduled for the first semester of 2015. 
Detailed chemical abundances for at least ten clusters in which six or more stars 
have been observed will be published in the second data release.

\section{Consistency\label{sec5}}

\begin{figure}
\center
\includegraphics[scale=0.4]{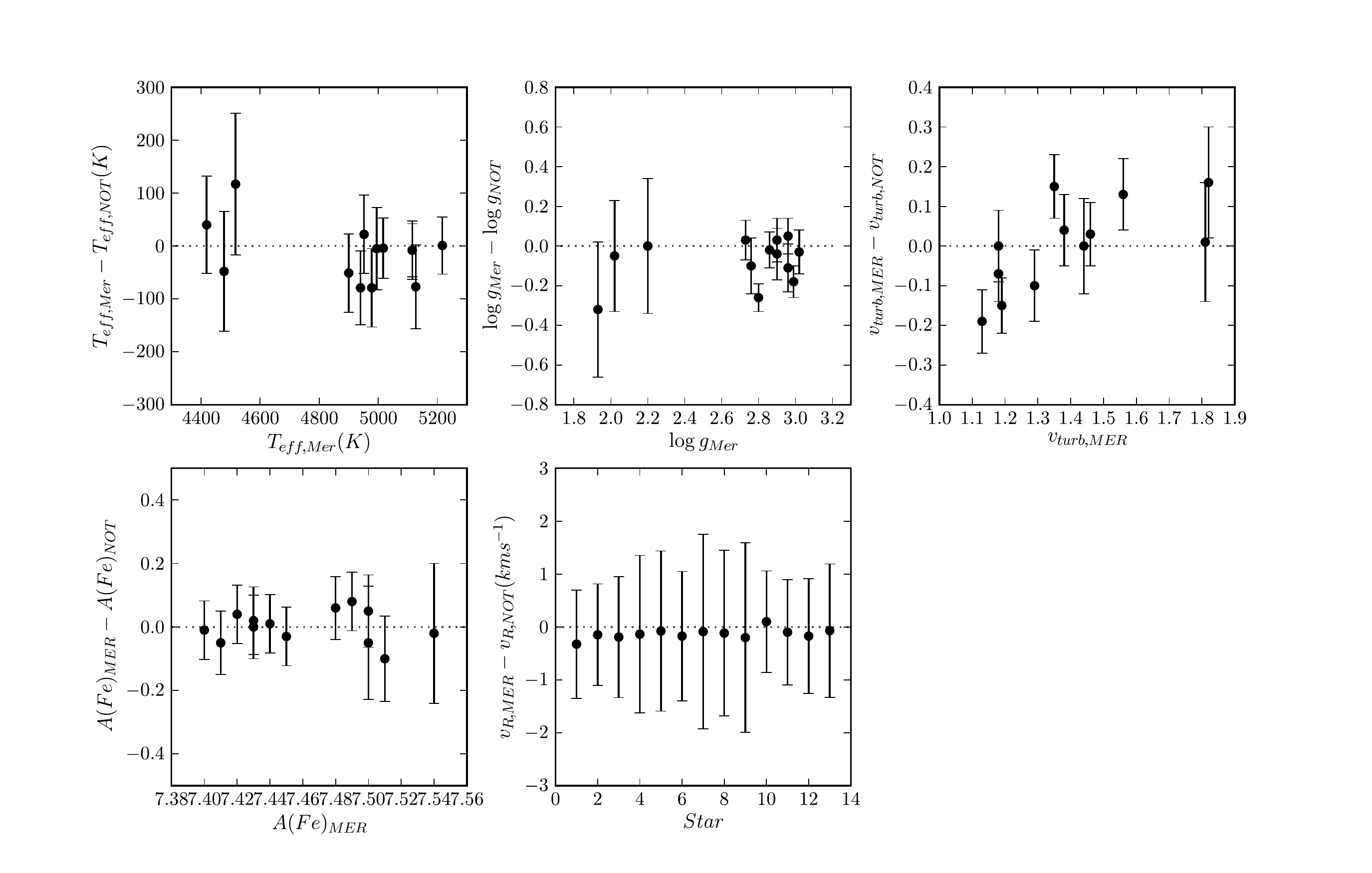}
\caption{\label{fig1} Comparison among stellar parameters derived for spectra acquired with FIES@NOT and HERMES@Mercator for the same stars using DAOSPEC+GALA approach.}
\end{figure}

One of the OCCASO requirements is the homogeneity between telescopes, method and model atmospheres
used, and in the same scale than the GES-UVES 
abundances. For this reason we are performing different tests. For example, to 
ensure the homogeneity among telescopes we have observed several stars in common 
in all of them. The preliminary results in the comparison between NOT and 
Mercator telescopes using the DAOSPEC+GALA approach are shown in  Fig.~\ref{fig1}. The values obtained for each 
telescope agree within the uncertainties. We are working on the other comparisons 
described above such as results obtained from different methods, different 
atmosphere models, wavelength ranges, etc.

\section{Summary and future work}

The OCCASO survey has been designed as the northern counterpart of the GES-UVES 
observations using the high-resolution spectroscopic facilities available at the Spanish 
observatories. OCCASO aims to derive abundances for a sample of 
20-25 OCs older than 0.5 Gyr. Together with the other 20-25 OCs in the same 
age range included in the Southern sample of GES-UVES it will constitute the largest homogeneous sample available until now, and that will be key in the study of the 
formation and evolution of the Galactic disc.

Up to November of 2014, OCASSO has completed 44 observing nights in which 524 
spectra for 87 stars belonging to 18 OCs have been acquired. The minimum requirement of 
at least 6 stars has been met for 11 clusters. The data reduction of all 
spectra acquired until now with the NOT and Mercator telescopes have been completed with 
the dedicated pipelines available for each instrument. The atmosphere 
parameters and chemical abundances for these stars have been obtained using the 
classical equivalent width approach as implemented by DAOSPEC and GALA. The data from the CAHA 2.2m telescope is being reduced with a new pipeline 
developed and kindly provided to us by J. Maiz-Apellaniz. Moreover we are 
working on deriving atmosphere parameters and abundances with spectral synthesis methods such as FERRE, MATISSE and iSpec.

One of the goals of OCCASO is the internal homogeneity, in addition to derive abundances in the same scale than the GES-UVES. 
To ensure this, we have observed several stars in common among 
telescopes and also with GES. We are performing exhaustive tests to achieve the level of homogeneity needed for the scientific goals.

The observing time awarded by OCCASO will finish in semester 15A. However, 
because of the bad weather in previous observing runs we will need more time at 
NOT to complete the faintest clusters in our sample. In any case, we plan to publish 
the first OCCASO data release in the first semester of 2015. It will include 
the atmosphere parameters and iron abundances for all the stars observed until 
now. In a second data release, expected for the second semester of 2015, we will 
publish the detailed abundances for at least ten OCs in which six or more stars have 
already been observed. Finally we are planning an extension of OCCASO to fainter clusters using 
HORS \cite{2014SPIE.9147E..8JP}, a new high-resolution spectrograph for the 10m 
GTC telescope expected to see first light in 2015.

\small  
%
\section*{Acknowledgments}   
%
This work was supported by the Spanish Ministry of Economy and Competitiveness - FEDER 
through grants AYA2012-39551-C02-01, AYA2010-16717 and ESP2013-48318-C2-1-R.
The work reported on in this publication has been fully or partially supported 
by the European Science Foundation (ESF), in the framework of the GREAT Research 
Networking Programme.%

%
\end{document}